# On two direct methods for measurement of interfacial tension at microdroplet surfaces

J. Tothova, M. Richterova and V. Lisy

*Institute of Physics, P.J. Safarik University, Jesenna 5, 041 54 Kosice, Slovakia*

**Abstract.** It has been shown by Yeung *et al.* [J. Colloid Interface Sci., 1998, **208**, 241] that in the presence of surfactants, the interfacial tension (IFT) of microscopic droplets differs significantly from IFT of macroscopic drops for the same surfactant solutions. As a result, IFT between two immiscible liquids can strongly depend on the experimental design. In the present work a simple theoretical description of two possible techniques for measurement of IFT at microdroplet surfaces is proposed: the spinning drop and the micropipette methods. We take into account that for microdroplets the IFT can be not the unique parameter determining the surface elastic energy. The previous interpretation of the experiments is generalized within the Helfrich's concept of interfacial elasticity. Simple equations are obtained that could be used to test the Helfrich's theory and in the direct determination, in addition to IFT, of such characteristics of the interface as the bending rigidity and spontaneous curvature.

## Introduction

Interfacial tension (IFT) is one of the most important physical properties of fluid-fluid interfaces [1]. In the absence of surfactants, IFT between two immiscible fluids is an intrinsic property which does not depend on the size or geometry of the interface. With the addition of surfactants it is, however, no longer true. It has been demonstrated both theoretically and experimentally that IFT depends on the partitioning of surfactants between the bulk phases and the interface [2]. The distribution of surfactants is determined by geometric factors such as the volume fractions and specific interfacial areas of the two fluids. As a result, the IFT of microscopic droplets may differ significantly from that of macroscopic drops for the same surfactant solutions [1, 2]. This difference is due to the effect of an enlarged interfacial area [2]. In this regard the necessity to conduct measurements of IFT on a length scale that is characteristic for the individual elements of the emulsion is evident. But when the size of the droplets dispersed in emulsion decreases, IFT becomes not the unique quantity determining the thermodynamic properties of the interface. The existence of stable spherical droplets in microemulsions cannot be explained using only the concept of IFT. The free energy of the droplet, in addition to the elastic term $\sigma A$ ($\sigma$ is the interfacial tension and $A$ is the surface area), must incorporate further contributions [3]. Within the widely accepted Helfrich's model of interfacial elasticity [4], the interface is characterized not only by $\sigma$ and the droplet radius $R$, but additional parameters are needed: the bending rigidity and the Gaussian (or saddle splay) modules $\kappa$ and $\bar{\kappa}$, and the spontaneous curvature, $C_s$.

The determination of these parameters has been attempted by a number of macroscopic and microscopic methods [5]. However, different experimental methods often yield very different values for the parameters of the surface film at similar conditions (for the discussion see [6]). It is thus of considerable interest to have available theoretical description of various experiments on emulsion droplets that could serve as alternative probes of the interfacial characteristics. One of such methods, the micropipette technique, originally developed for the biophysical sciences [7 – 9], has been recently introduced to directly measure IFT of micrometer-sized droplets [2, 10]. The other technique, the spinning drop method, is known for years [11] and has been very successfully applied in examination of ultralow interfacial tensions down to $10^{-6}$ mN/m [1, 12 – 14].

In the present paper we propose descriptions of the two methods that generalize the previous interpretations to the case of small droplets, when the Helfrich's curvature energy can play a role in the droplet formation and stability. Simple formulas are obtained that relate the phenomenological characteristics of the interface to the quantities observed in the experiments. It will be shown that the modifications of the previous equations used in the interpretation of the measurements can be in some cases significant even for relatively large droplets, such as in macroemulsions. In spite of experimental difficulties expected in the measurements on very small droplets, we believe that the described techniques in combination with other methods such as videomicroscopy could serve as direct methods for the determination of important characteristics of the interface. In particular, the Helfrich's theory could be tested in a direct way.

## The Laplace-Helfrich equation

For the following we need a generalization of the well-known Laplace equation describing the pressure difference across the curved interface [15],

$$\Delta p = \sigma \left( \frac{1}{R_1} + \frac{1}{R_2} \right), \qquad (1)$$

where $R_1$ and $R_2$ are the radii of curvature. Within the Helfrich's theory, the equilibrium shape of a membrane that covers the vesicle or emulsion droplet is determined by the minimization of the Helmholtz free energy (or the "shape energy" [3, 4]), which may be written as

$$E_S = -\Delta p V + \sigma A + \int dA \left[ \frac{\kappa}{2} \left( \frac{1}{R_1} + \frac{1}{R_2} - \frac{2}{R_s} \right)^2 + \frac{\bar{\kappa}}{R_1 R_2} \right]. \qquad (2)$$

Here $\Delta p = p_1 - p_2$ is the pressure inside the droplet minus outside, $V$ is the volume of the droplet, $R_s$ is the spontaneous radius of curvature, and the parameters $\kappa$ and $\bar{\kappa}$ are described in Introduction. The spontaneous curvature $C_s$ describes the effect of a possible asymmetry of the interface or its environment [16]. The minimization of Eq. (2) results in the equilibrium condition [16, 17]

$$-(p_1 - p_2) + 2\sigma H + \kappa(C_s - 2H)(2H^2 - 2K + C_s H) - 2\kappa \Delta H = 0, \qquad (3)$$

where $H = (1/R_1 + 1/R_2)/2$ is the mean and $K = 1/R_1R_2$ the Gaussian curvature, and $\Delta$ is the Laplace-Beltrami operator on the surface [16]. This shape equation generalizes the Laplace condition (1) and represents the balance of normal forces per unit area. In general, it contains the complicated stresses of curvature elasticity so that it is difficult to write it explicitly. Below we shall consider only the cases of spherical and cylindrical surfaces, when Eq. (3) takes a simple form. The curvature radii of these surfaces are assumed to be much larger than the thickness of the surface layer, which is the condition of applicability of the used phenomenology.

**The spinning drop method**

In the spinning drop tensiometry a fluid drop of mass density $\rho_1$ is suspended in a liquid with the density $\rho_2 > \rho_1$. The drop and the liquid are contained in a horizontal tube cell. The cell rotates about its longitudinal axis. At low rotational velocities $\omega$, the drop will take on an ellipsoidal shape, but when $\omega$ is sufficiently large, it will become cylindrical. Under this latter condition, the radius $r$ of the cylinder is determined by the interfacial tension, the density difference, and the rotational velocity. The Vonnegut's equation [11]

$$\sigma = \frac{1}{4}(\rho_2 - \rho_1)\omega^2 r^3 \qquad (4)$$

can be derived in various ways [11, 18], for example, by considering the change of the kinetic energy $(1/2)\int \rho v^2 dV$ of the system drop plus liquid with respect to the case when the cell contains only the liquid. The velocity in the stationary limit is simply determined by $\omega d$ ($d$ is the distance from the axis of rotation), so that the change of the kinetic energy due to the presence of the lighter cylindrical drop is

$$E_K = \frac{\pi}{4} L \omega^2 r^2 (\rho_2 - \rho_1), \qquad (5)$$

where $L$ is the length of the cylinder, assumed to be much larger than $r$. The total energy change is obtained by adding the surface energy of the cylinder, $2\pi L r \sigma$. The minimization of the total energy change with respect to $r$ (at constant volume of the droplet) then yields Eq. (4). In our case the Helfrich's corrections to the surface energy are taken into account. The principal curvature radii of the cylinder are $R_1 = r$ and $R_2 = \infty$. Then the surface energy is

$$F_S = 2\pi L r \sigma + \pi L r \kappa \left(\frac{1}{r} - \frac{2}{R_s}\right)^2. \qquad (6)$$

The minimization of the total energy change $F_S + E_K$ at constant $V$ gives

$$\sigma = \frac{1}{4}(\rho_2 - \rho_1)\omega^2 r^3 - \frac{\kappa}{r^2}\left(\frac{3}{2} - 4\frac{r}{R_s} + 2\frac{r^2}{R_s^2}\right). \qquad (7)$$

The quantity $\alpha = \sigma + 2\kappa/R_s^2$ which appears in this equation is regarded as the surface tension for the plane interface [3]. If we assume an initial spherical drop with the radius $R$, and the final cylindrical drop with $L \gg r$, then the leading correction to the Vonnegut's equation is given by the second term on the right hand side of Eq. (7),

$$\alpha \approx \frac{1}{4}(\rho_2 - \rho_1)\omega^2 r^3 - \frac{3\kappa}{2r^2}. \qquad (8)$$

It was assumed here that $R_s > R$ as usually in microemulsions, so that the terms $\sim r/R_s$ and $(r/R_s)^2$ are small. In general, however, $R_s$ is an unknown phenomenological parameter that can vary for different systems and Eq. (7) has to be used. The bending rigidity $\kappa$ also varies from system to system and very different values for it can be found in the literature, changing in some two orders in magnitude [6]. In usual spinning drop measurements the volume of the drop is large, usually $\sim 0.1$ cm$^3$, so that the corrections to the Vonnegut's expression are very small ($\kappa$ is much smaller than the lowest detectable $\sigma \sim 10^{-6}$ mN/m in this technique). When the size of the droplet decreases, these corrections become play a role, and the measurements possess not $\sigma$ but some effective (apparent) surface tension $\sigma_{eff} = (\rho_2 - \rho_1)\omega^2 r^3/4$. The difference between $\sigma$ and $\sigma_{eff}$ can be considerable, especially for low IFT. So, for a cylindrical droplet with $r \sim 10$ $\mu$m or smaller and a typical $\kappa \sim 10^{-19}$ J or larger, $\kappa/r^2$ becomes larger than about $3 \cdot 10^{-5}$ mN/m; such or even smaller values of IFT are often encountered in emulsion research. A question arises about the range of possible $\sigma_{eff}$ detectable by the current spinning drop devices. A simple estimation shows that for a droplet with $r \sim 10$ $\mu$m and the density difference $\rho_2 - \rho_1 \sim 3 \cdot 10^2$ kgm$^{-3}$ (the density of water minus a typical density of aliphatic oils like hexane, octane,…), we find for $\omega = 2\pi f \sim 5\pi \cdot 10^2$ s$^{-1}$ the value $\sigma_{eff} \sim 3 \cdot 10^{-4}$ mN/m ($f \sim 250$ s$^{-1}$ was taken as an upper limit for the rotational velocities of current devices). Consequently, varying $\omega$, $\sigma_{eff}$ could be measured in a rather broad range of magnitude, if $r \sim 10$ $\mu$m. Alternatively, smaller droplets could be examined for sufficiently high $\omega$. In combination with videomicroscopy such experiments seem to be realistic (note that using videomicroscopy even the shape fluctuations of microdroplets are successfully studied [20]).

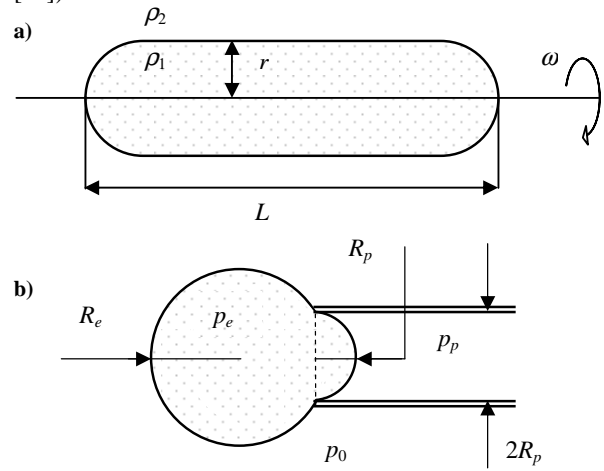

**Figure:** Schematic illustrations of the spinning drop method (a) and the micropipette technique (b).

**Micropipette technique**

In this microtensiometry technique, the droplet is first captured at the tip of the glass micropipette and then sucked into the pipette [1, 2, 10]. The IFT is calculated from the minimum pressure at which the droplet extends a hemispherical protrusion into the pipette, and by using the Laplace equation (1). If $R_p$ is the inner radius of the pipette, $R_e$ is the radius of the exterior spherical segment of the droplet, $p_0$ is the pressure outside this segment and $p_p$ inside the pipette, see Figure, Eq. (1) yields

$$p_0 - p_p = 2\sigma\left(\frac{1}{R_p} - \frac{1}{R_e}\right). \tag{9}$$

If we want to take into account the Helfrich's bending energy of the interface, Eq. (2), we have to use the Laplace-Helfrich equation. For the pressure difference across the spherical surface, Eq. (3) simplifies to

$$\Delta p = \frac{2\sigma}{R} + \frac{4\kappa}{RR_s}\left(\frac{1}{R_s} - \frac{1}{R}\right), \tag{10}$$

where $R$ is the radius of the surface. Applying Eq. (10) to the external segment of the droplet with the pressure $p_e$ inside it,

$$p_e - p_0 = \frac{2\sigma}{R_e} + \frac{4\kappa}{R_e R_s}\left(\frac{1}{R_s} - \frac{1}{R_e}\right),$$

then to the aspirated hemisphere,

$$p_e - p_p = \frac{2\sigma}{R_p} + \frac{4\kappa}{R_p R_s}\left(\frac{1}{R_s} - \frac{1}{R_p}\right),$$

and subtracting the two equations, we obtain

$$p_0 - p_p = 2\sigma_{eff}\left(\frac{1}{R_p} - \frac{1}{R_e}\right), \tag{11}$$

where the apparent IFT is now

$$\sigma_{eff} = \sigma + \frac{2\kappa}{R_s}\left(\frac{1}{R_s} - \frac{1}{R_p} - \frac{1}{R_e}\right). \tag{12}$$

Thus, if the IFT is known for the plane interface, this method could be used, in combination with other methods, for the determination of $\kappa$ and $R_s$. The micropipette experiments were first designed for the study of vesicles. Note that there is a difference between our description and the interpretation of the micropipette experiments by Evans and Rawicz [9] carried out on vesicles. In Ref. [9] the observed $\sigma_{eff}$ was related to the change of the surface area at the droplet deformation. This was done using the theory of small fluctuations in the shape of spheroidal droplets exhibiting a constant excess area of the surface. In the interpretation [9] this excess area is at small tensions identified with the change of the surface area $\Delta A$ during the droplet deformation, and the apparent tension corresponds to our quantity $\sigma$. At high tensions the dependence $\sigma \sim \Delta A$ is postulated in agreement with the observations. The spontaneous radius of curvature was taken to be infinite, which is true only for free vesicles. Our approach is essentially different since we assume $\sigma$ to be an intrinsic property of the droplet and distinguish it from the apparent $\sigma_{eff}$ that is expressed through the parameters of the surface layer $R_s$, $\sigma$, and $\kappa$. As to the thermal fluctuations of the droplet shape, we do not take them into account neither in the description of the micropipette experiments nor in the spinning drop method. In both the methods relatively large extensions of the droplets take place. Then the undulations of the surface are suppressed [9], so that their influence on the measured quantities is expected to be small. Using our formula (12) together with simple geometrical consideration allow one to extract the quantity $\kappa/R_s$ from the region of the linear dependence of $\sigma_{eff}$ on $\Delta A$.

## Conclusion

The present work comes from two assumptions: 1) The observed interfacial tension should depend on the size of investigated emulsion droplets. 2) At least for small (micrometer-sized or smaller) droplets, IFT is not the unique phenomenological parameter characterizing the elastic energy of the droplet surface. The first assumption is a particular consequence of the findings by Yeung *et al.* (2), who have shown that, in the presence of surfactants, the IFT between two immiscible liquids can strongly depend on the experimental design. As to the second claim, it is well known that the existence, stability, and other properties of microemulsion droplets cannot be understood on the basis of the surface elastic energy alone [3]. The widely accepted resolution of this problem proposed by Helfrich [4] consists in the introduction of the additional (bending or curvature) energy of the interface.

Within the Helfrich's concept of interfacial elasticity, we generalize the description of two methods of the measurements of IFT: the spinning drop technique known since 1942 [11], and the micropipette method, recently introduced in emulsion research in the work [2]. The essential advantage of these techniques, if compared with other (indirect and complicated in interpretation) methods like the scattering of light and neutrons [6, 12], is their simplicity. For both the methods new simple equations are obtained, which relate the observable quantities to the surface characteristics of the droplets. These characteristics besides the IFT include the bending rigidity and the spontaneous curvature of the interface. In favorable conditions, when the IFT is small, they could be directly extracted from the experiments. The described methods thus could serve as direct techniques for the determination of the most important parameters in the study of fluid-fluid interfaces, and for testing the Helfrich's theory.